# Validation of the GATE Monte Carlo simulation platform for modelling a CsI(Tl) scintillation camera dedicated to small animal imaging

**D Lazaro[1], I Buvat[2], G Loudos[3,4], D Strul[5], G Santin[6], N Giokaris[3,4], D Donnarieix[1,7], L Maigne[1], V Spanoudaki[3,4], S Styliaris[3,4], S Staelens[8] and V Breton[1,9]**

[1] Laboratoire de Physique Corpusculaire, CNRS/IN2P3, Université de Clermont-Ferrand, 24 avenue des Landais, 63177 Aubière Cedex 1, France
e-mail : lazaro@clermont.in2p3.fr
[2] INSERM U494, CHU Pitié-Salpêtrière, 91 Boulevard de l'Hôpital, 75634 Paris Cedex 13, France
[3] Institute of Accelerating Systems and Applications of Athens, P.O. Box 17214, 10024, Athens, Greece
[4] University of Athens, Physics Department, Nuclear and Particle Physics Division, Panepistimioupolis Ilisia, 157 71 Athens, Greece
[5] Institute of High Energy Physics, University of Lausanne, CH-1015 Dorigny, Switzerland
[6] ESA-ESTEC, Noordwijk, The Netherlands
[7] Département de Curiethérapie-Radiothérapie, Centre Jean Perrin, 63000 Clermont-Ferrand, France
[8] ELIS Department, Ghent University, Sint-Pietersnieuwstraat 41 B-9000 Ghent, Belgium

E-mail: lazaro@clermont.in2p3.fr

**Abstract**.

Monte Carlo simulations are increasingly used in scintigraphic imaging to model imaging systems and to develop and assess tomographic reconstruction algorithms and correction methods for improved image quantitation. GATE (GEANT 4 Application for Tomographic Emission) is a new Monte Carlo simulation platform based on GEANT4 dedicated to nuclear imaging applications. This paper describes the GATE simulation of a prototype of scintillation camera dedicated to small animal imaging and consisting of a CsI(Tl) crystal array coupled to a position sensitive photomultiplier tube. The relevance of GATE to model the camera prototype was assessed by comparing simulated $^{99m}$Tc point spread functions, energy spectra, sensitivities, scatter fractions and image of a capillary phantom with the corresponding experimental measurements. Results showed an excellent agreement between simulated and experimental data: experimental spatial resolutions were predicted with an error less than 100 $\mu$m. The difference between experimental and simulated system sensitivities for different source-to-collimator distances was within 2%. Simulated and experimental scatter fractions in a [98-182 keV] energy window differed by less than 2% for sources located in water. Simulated and experimental energy spectra agreed very well between 40 and 180 keV. These results demonstrate the ability and flexibility of GATE for simulating original detector designs. The main weakness of GATE concerns the long computation time it requires: this issue is currently under investigation by the GEANT4 and the GATE collaborations.

## 1. Introduction

Monte Carlo simulations are extensively used nowadays to address various issues related to scintigraphic imaging, for instance to design and optimize imaging systems and collimator or to develop and assess correction methods and tomographic reconstruction algorithms for improved image quantitation (Zaidi 1999). Two categories of Monte Carlo codes are currently available in nuclear imaging (Zaidi 1999, Buvat and Castiglioni 2002): general purpose codes (EGS4, GEANT, MCNP, ITS), developed for high-energy physics or dosimetry, and dedicated codes (e.g., SIMIND, SimSET, SimSPECT, PETSIM), especially designed for Single Photon Emission Computed Tomography (SPECT) and/or Positron Emission Tomography (PET). Although dedicated codes are usually convenient to use and well suited to the simulations of commonly used SPECT and PET configurations, most of them do not accurately simulate all the system components, such as the collimator or the components located behind the crystal, while these components can substantially affect the final image characteristics (De Vries *et al* 1990, Yanch and Dobrzeniecki 1993). Also, dedicated codes have limited flexibility for simulating non-conventional imaging device. For such applications, general purpose codes might be preferable. GATE (GEANT4 Application for Tomographic Emission) (Santin *et al* 2003) is a generic simulation platform based on a general purpose code GEANT4 and designed to answer the specific needs of PET/SPECT applications. Several research institutes dealing with SPECT and PET are involved in the development and validation of GATE within the OpenGATE Collaboration (http://www-lphe.epfl.ch/~PET/research/gate). In addition to the many potentialities provided by GEANT4, GATE includes specific modules necessary to perform realistic SPECT and PET simulations, including modules managing time and time-dependent processes (detector and source movements, radioactive decay, dynamic acquisitions), complex and voxel-based source distributions and easy description of scanner geometry.

This paper presents the use and validation of GATE for simulating acquisitions performed on a scintillation camera prototype dedicated to small animal imaging built and tested at the Institute of Accelerating Systems and Applications of Athens (IASA). Small field-of-view (FOV) scintillation cameras based on position-sensitive photomultiplier tubes (PSPMTs) have been developed during the past decade and have already demonstrated their suitability for small-organ imaging (Zaidi 1996), radiopharmaceutical testing (Loudos *et al* 2003) or scintimammography (Maublant *et al* 1996, Scopinaro *et al* 1999, Williams *et al* 2000). Characteristics and performances of such devices have been well described (Vittori *et al* 1998, Pani *et al* 1997): a spatial resolution of about 2 mm can be achieved at the expense of the energy resolution, which is about 30%. Few papers report complete simulations of small field-of-view scintillation cameras based on a crystal array (Vittori *et al* 2000, Garibaldi *et al* 2001). Such simulations should help optimize the geometry and components of the imaging device, test and assess imaging and processing strategies.

Section 2 briefly describes the main features of the GATE simulation platform, the experimental scintillation camera prototype, as well as the scintillation camera model developed with GATE and the validation procedures used to assess the relevance of the simulated data. In section 3, experimental energy spectra, point spread functions (PSF), sensitivity, scatter fractions and the image of a capillary phantom are compared with simulation results for validation. A discussion of the results and of GATE features concludes the manuscript.

## 2. Material and methods

### 2.1. GATE simulation platform

The GATE simulation platform is based on the GEANT4 toolkit (Santin *et al* 2003). GEANT is a simulation code that has been developed at CERN for more than 2 decades for the description of interactions between particles and matter. The first version of the code, GEANT3, was written in Fortran and has been extensively used around the world in high-energy physics and for medical applications (e.g., Tsang *et al* 1995, Rogers and Gumplinger 1999, Berthot *et al* 2000, Porras *et al* 2002). A new version of the code, GEANT4 (RD 44 Collaboration 1998) is developed since 1994 and is written in the C++ object-oriented language. In this new version, the modeling of electromagnetic physical processes has been extended to reliably cover electromagnetic interactions from 250 eV to 100 GeV (Apostolakis *et al* 1999). This modeling makes use of recently updated libraries developed at the Lawrence Livermore National Laboratory (EADL, EEDL and EPDL97).

As the global architecture of GATE has been described in details (Santin *et al* 2003, Strul *et al* 2003), we only underline here the main features of GATE. On top of GEANT4, GATE includes specific modules that have been developed to meet specific requirements encountered in SPECT and PET and to facilitate the use of the code. A user-friendly mechanism based on scripts is used to easily define all the simulation parameters, including complex detector, or phantom and source distribution geometry modeling. GATE can model time-dependent processes, through the use of a virtual clock (Santin *et al* 2003), which allows the management of scanner or patient movements and tracer kinetics for instance.

In this work, GATE was used to simulate the IASA scintillation camera (see section 2.2).

### 2.2. The IASA scintillation camera

*2.2.1. Description of the prototype.* The IASA scintillation camera prototype consists of a 3 mm thick scintillating CsI(Tl) array coupled to a PSPMT Hamamatsu R2486 (Malatesta et al 1998). The crystal array is 4.6 cm in diameter and is made of square pixels, covering the whole circular section of the crystal. Each pixel is separated from the others by a 250 $\mu$m-thick diffusive white layer (epoxy). The maximum number of pixels along the diameter of the crystal array is 41 and the arrangement of the pixels is shown in figure 1. A detailed description of the CsI(Tl) optical and mechanical characteristics can be found elsewhere (Vittori et al 1998).

The PSPMT is equipped with two resistive chains connecting 8 + 8 crossed anode wires. The readout of the 16 anode signals enables calculation of the centre of gravity (COG) of the electron cloud and the subsequent determination of the exact position of the incident photon in the (X,Y) plane. To avoid edge effects, anode wires that carry less than 5 % of the total anode signal are disregarded in the COG calculation.

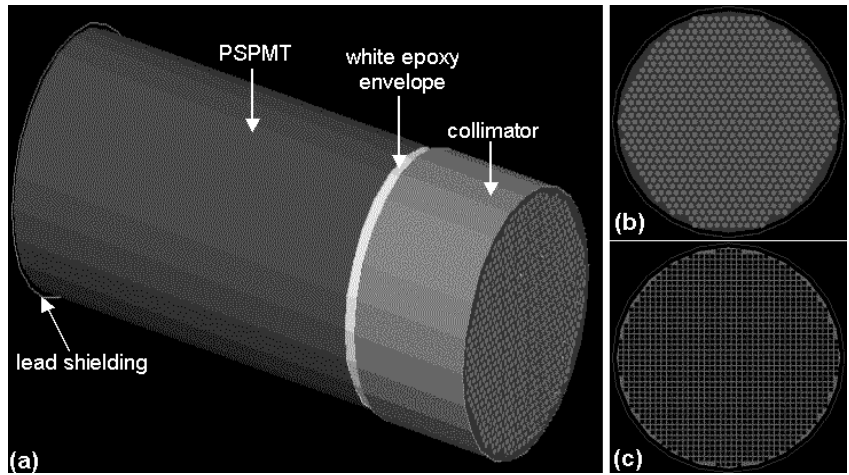

**Figure 1.** (a) Picture of the IASA scintillation camera simulated with GATE: the different components of the detector head are represented (collimator, white epoxy envelope surrounding the crystal, PSPMT and shielding); (b) the hexagonal hole collimator geometry; (c) the CsI(Tl) crystal array geometry.

The prototype is equipped with a removable low energy high resolution collimator (2.75 cm height, 0.4625 mm septal thickness) with a 1.12 mm flat-to-flat distance of the hexagonal parallel holes. The whole detection head is surrounded with a 5 mm thick lead shielding.
Figure 1 shows a scheme of the experimental setup obtained by simulation, as well as the details of the collimator and CsI(Tl) crystal array. More details about this prototype can be found in (Loudos *et al* 2003).

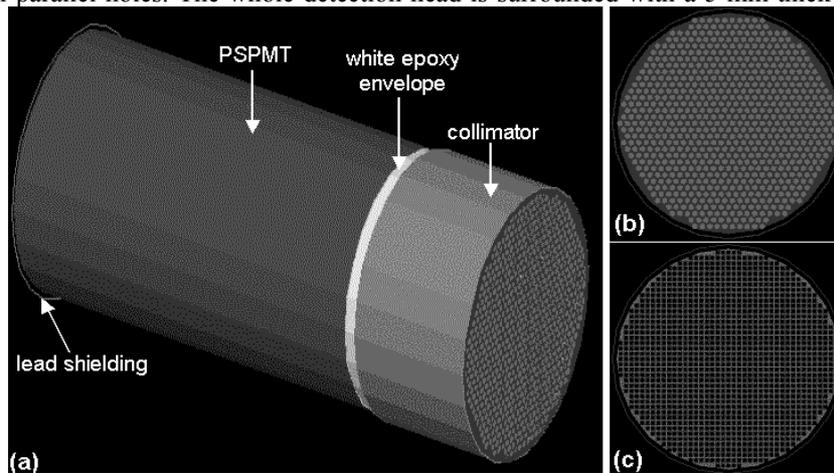

*2.2.2. Monte Carlo model of the IASA camera.*             To simulate the IASA camera, the collimator, the pixelated crystal (diffusive white layer envelope and crystal array) and the lead shielding were modelled. As the pixelated crystal is 3 mm thick, about 30% of the 140 keV incident photons leave the crystal without interacting. Backscatter on the PSPMT glass entrance window is taken into account by modelling the PSPMT by a single Plexiglas® layer, 8 cm thick and 7.6 cm in diameter (De Vries *et al* 1990). A picture of the IASA scintillation camera simulated with GATE (based on GEANT4 version 4.4.0 for this study) is shown in figure 1.

In all simulations, each gamma ray was tracked through the object and detector until its energy fell below the energy cut or the gamma ray escaped from the system, defined by the user as a large volume including the whole detection device (collimator, crystal array and PSPMT) and the imaged object (source and/or phantom). The energy cut was set to 10 keV, which is accurate enough as the experimental device provided the energy spectrum only down to 40 keV. X-rays resulting from fluorescence and Auger processes were tracked to ensure accurate modelling of interactions within the collimator of the camera, whereas secondary electrons were not tracked to speed up the simulation.

The physical processes involving gamma interactions (photoelectric effect, Compton scattering and Rayleigh scattering) were modelled using the electromagnetic low-energy package of GEANT4. If energy was deposited in the crystal at more than one site, the centroid of all interaction points weighted by the ratio of the deposited energy to the total energy deposited in the crystal was calculated to deduce the location of the event. All deposited energies were summed to associate a single energy value to the detected event. Spatial coordinates of the centroid and associated energy were stored. Optical photon tracking was not modelled to save computation time.

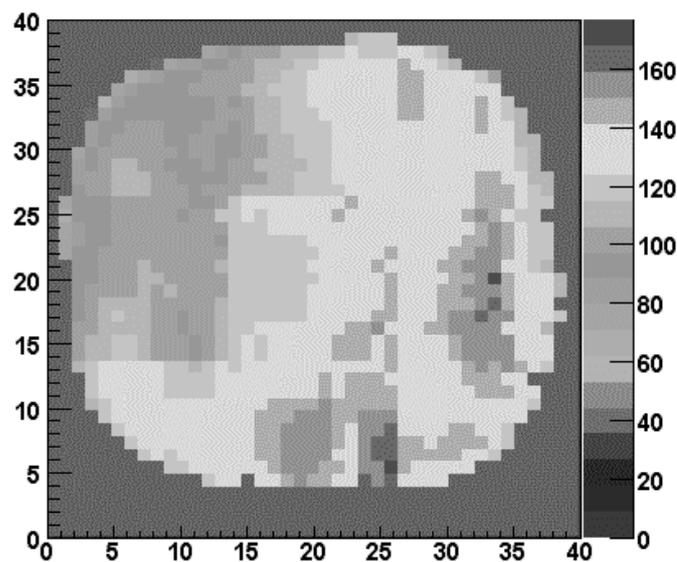

**Figure 2**. Map of the Full Energy Peak (FEP) channel values.

As already reported in (Vittori *et al* 1998) and (Malatesta *et al* 1998), the pixels of the crystal array may present large light yield differences, which worsens the energy resolution of the system. To accurately reproduce the measured energy resolution, this effect had to be taken into account in the simulations. For doing so, the energy response of the pixelated crystal was experimentally measured by irradiating the detection area of CsI(Tl) with a 7.4 x 10[8] Bq liquid source of [99m]Tc (140 keV) contained in a capillary of 1.3 mm external diameter and 4 mm length and located at a few millimetres from the crystal surface. The detection area of the crystal was subdivided into a matrix of 41 x 41 pixels and the energy spectra of the events detected within each pixel were stored. The Full Energy Peak (FEP), defined as the channel in which the maximum value of the photopeak was recorded, as well as the full width at half maximum (FWHM) of the local photopeak, were computed for each pixel.

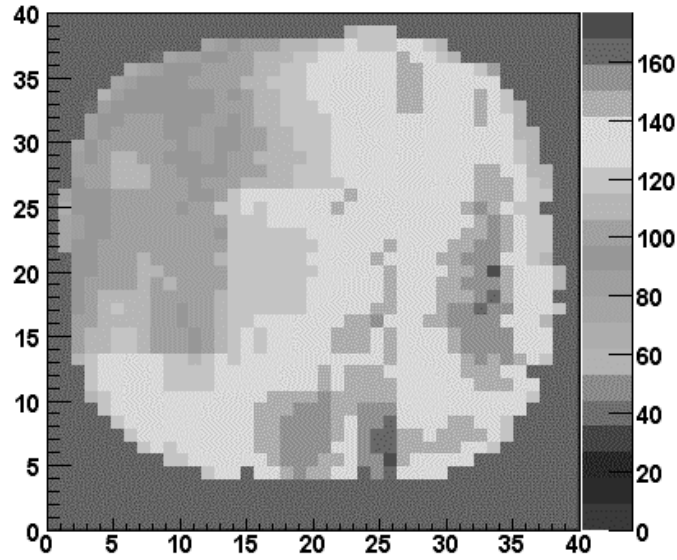

Figure **2** represents the 41 x 41 map of FEP values: the FEP channel varied from 80 to 170 keV from one pixel to another.

This map was used to model the light yield differences in the simulation: depending on the interaction point of the photon in the crystal array, the pixel where the interaction occurred was determined and the associated energy deposit was shifted to the value predicted by the FEP map for this pixel.

Using GEANT4, the simulation of the complete detection process is possible including the modelling of the scintillation process and of the PSPMT. However, using the Monte Carlo approach to model these effects is very time consuming. We therefore used analytical convolution models to account for the intrinsic energy and spatial responses of the PSPMT and subsequent electronics, using a Gaussian energy response function and a Gaussian point spread function. As the FWHM of the photopeak results both from the broadening of the photopeak introduced by the large light yield differences between adjacent pixels in the crystal and from the energy blurring caused by statistical fluctuations occurring within the crystal, the PSPMT and electronics, the energy response function was modelled in two steps: (1) first, the value of deposited energy was shifted using the FEP map as described above, (2) the resulting energy was blurred using a Gaussian function whose FHWM was given by the mean of the FWHM values (20.5% at 140 keV) calculated over all pixels of the crystal array. The energy resolution of this Gaussian energy response function was considered to be energy dependent following the relation (Knoll 1989):

$$FWHM(E) = \sqrt{\alpha + \beta E} \big/ E$$

$\alpha$ and $\beta$ were estimated by considering the 20.5% energy resolution measured at 140 keV and the energy resolution measured at 70 keV with $^{201}$Tl, yielding $\alpha$ = -8.46 x $10^6$ and $\beta$ = 1.76 x $10^5$ for FWHM(E) and E expressed in keV.

An intrinsic spatial resolution of $FWHM_{intrinsic}$ = 1.4 mm was previously measured in planar imaging (Loudos *et al* 2003) at IASA over the whole FOV and the spatial resolution $FWHM_{crystal}$ of the CsI(Tl) crystal array was determined to be 0.6 mm (Vittori *et al* 1998). The spatial resolution corresponding to the PSPMT and associated electronics was thus assumed to be 1.26 mm ($\sqrt{FWHM^2_{intrinsic} - FWHM^2_{crystal}}$). This value was higher than the spatial resolution given by the manufacturer (0.5 mm) but accounts for spatial distortions introduced by the anode wires on the edges.

### 2.3. Validation of GATE for the IASA camera

*2.3.1. Spatial resolution.* The spatial resolution of the IASA camera was characterized by the point spread functions in air at 2, 5 and 10 cm from the collimator. These were experimentally measured with a $^{99m}$Tc point source consisting of a 1.3 mm diameter capillary of 2 mm length (activity of 4.92 x $10^5$ Bq). A first set of measurements was performed with the source at the centre of the FOV, and a second set was performed with the source 1 cm off-centred. Images corresponding to the 40-180 keV energy windows were obtained. A 10 mm thick profile through the point source was drawn for each of the six images (two source locations and three source-collimator distances) and the FWHM values were calculated.

The same six configurations were simulated using GATE. Monoenergetic gamma rays (140 keV) were emitted in the angle of acceptance of the scintillation camera in order to decrease computation time. The initial activity of the source was set to 4.92 x $10^5$ Bq and about 140 million photon histories were generated (CPU time of about 8 h on a biprocessor Pentium III 1GHz computer). Events were collected between 40 and 180 keV and associated images were calculated. Profiles through these images yielded the FWHM to be compared with the experimental values.

Point spread functions were also measured for the $^{99m}$Tc point source in water. The source contained an initial activity of 3.54 x $10^5$ Bq in a 1.3 mm diameter capillary of 3 mm length and was located at the centre of the FOV, at 12 cm from the collimator, under a cylindrical phantom of 8 cm diameter and 11 cm length filled with 4 or 10 cm of water. Images corresponding to the 40-180 keV energy windows were obtained and 10 mm thick profiles centred on the line source were drawn. The same configurations were simulated with GATE: the source activity considered in the simulations was the same as the experimental value and about 415 million photon histories were generated in $4\pi$ sr (CPU time of about 24 h).

*2.3.2. Sensitivity.* The system sensitivity, defined as counts per second per kBq, was experimentally measured and calculated using GATE for a $^{99m}$Tc point source in air contained in a 1.3 mm diameter capillary of 3 mm length (initial activity of 4.47 x $10^5$ Bq), located at the centre of the FOV and at 2, 5, 10, 15 and 20 cm from the collimator. About 280 million photon histories were generated in $4\pi$ sr with GATE (CPU time of about 10 h). The radioactive decay was taken into account in the simulation.

*2.3.3. Scatter fraction.* The scatter fraction was defined as the ratio of the number of events in the image which come from photons scattered at least once in the phantom or in the collimator to the number of events coming from primary photons not scattered in the phantom.

Scatter fractions were evaluated experimentally for a thickness of water of 4 cm and 10 cm, for the experimental configurations previously described in section 2.3.1. Scatter fractions were estimated as described by (Manglos *et al* 1987): images of the point source were first obtained with the phantom empty ('in air' data) and then filled with a thickness of water of 4 cm and 10 cm ('in water' data). Scatter fraction SF was estimated as:

$$SF = \frac{\sum_i \left( water(i) - k.air(i) \right)}{\sum_i k.air(i)}$$

where water(i) and air(i) represent the number of counts in pixel *(i)* of the planar image in water and air respectively. The factor *k* is given by:

$$k = e^{-\mu d}$$

where $\mu$ is the attenuation coefficient of water at 140 keV (0.154 cm$^{-1}$) and d is the depth of the point source in water.

Scatter fraction values were experimentally calculated for 3 energy windows: 140 ± 25% keV (105–175 keV), 140 ± 30% keV (98–182 keV) and 140 ± 35% keV (91–189 keV), and corresponding values were derived from the simulations. Large energy windows were considered because of the poor energy resolution of the scintillation camera, so that all primary counts were included.

*2.3.4. Energy spectra.* The energy spectra were experimentally measured over the whole FOV: (1) in air, with the $^{99m}$Tc point source (activity of 4.92 x 10$^5$ Bq) located at the centre of the FOV, at 2 cm from the collimator, (2) in water, with the $^{99m}$Tc point source (3.54 x 10$^5$ Bq), located at the centre of the FOV at 12 cm from the collimator, under the cylindrical phantom filled with 4 cm and 10 cm of water.

The radioactive background coming from the equipment in the room was measured without source and then subtracted from the measured energy spectra as the background radioactivity was not simulated with GATE.

Simulations of different scintillation camera designs were first performed to characterize scatter within the device as a function of the design, for the source in air at 2 cm from the collimator. The role of the energy deposit within the white epoxy gap between the crystal pixels was studied by varying the gap: 250 $\mu$m (design 2) and 100 $\mu$m (design 3), 250 $\mu$m being the actual value of the interpixel size. In these two simulations, the PSPMT was not modelled. Then, the PSPMT was modelled (design 1) using a single 8 cm thick Plexiglas® layer, 7.6 cm in diameter. These dimensions were calculated from the weight of the PSPMT and its attenuating and scattering properties. For this design, the epoxy gap was 250 $\mu$m. The energy spectra obtained for these three designs were compared to the experimental spectrum obtained with the source in air.

For the source in water, the same configurations as described in section 2.3.1. were simulated.

*2.3.5. Image of a capillary phantom.* A phantom consisting in 5 parallel capillaries (1.55 mm in diameter, 6 cm in length), with a capillary-to-capillary distance of 3.5 mm was used. The capillaries were filled with $^{99m}$Tc solutions of different activities as shown in figure 3. The phantom was located at 1.5 mm from the collimator of the scintillation camera.

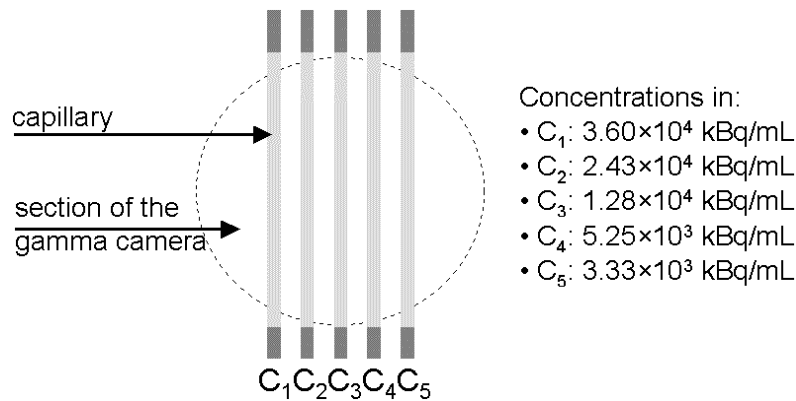

Concentrations in:
• $C_1$: 3.60×10$^4$ kBq/mL
• $C_2$: 2.43×10$^4$ kBq/mL
• $C_3$: 1.28×10$^4$ kBq/mL
• $C_4$: 5.25×10$^3$ kBq/mL
• $C_5$: 3.33×10$^3$ kBq/mL

**Figure 3**. Geometry and characteristics of the capillary phantom.

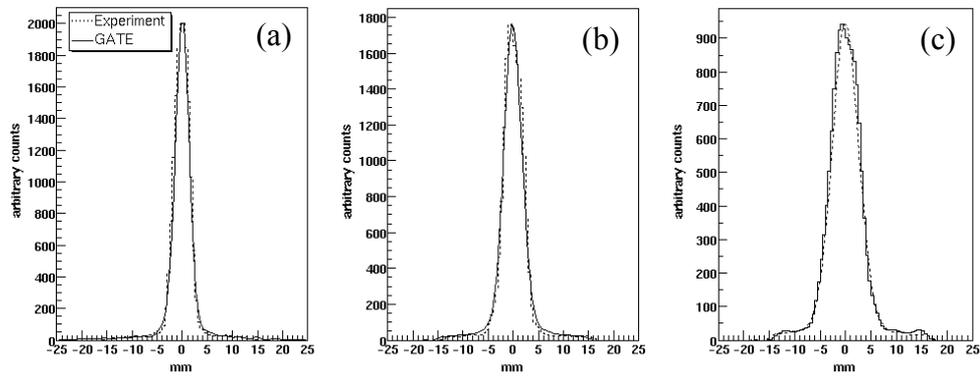

**Figure 4.** Measured and GATE simulated point spread functions for a centred $^{99m}$Tc point source located at (a) 2cm ,(b) 5 cm and (c) 10 cm from the camera.

The image of the phantom corresponding to the 40-180 keV energy window over the whole FOV was obtained and compared to the image resulting from the corresponding simulation. About 140 000 photons were experimentally detected within the 40-180 keV energy window. About one billion photon histories were generated in $4\pi$ sr in the simulation and about 30 000 photons were detected (CPU time of about 24 h).

## 3. Results

### 3.1. Spatial resolution

Simulated and experimental point spread functions are shown for the centred and off-centred sources in figure 4 and figure 5 respectively, where the experimental and simulated point spread functions were normalized to the same maximum.

Figure 5. Measured and GATE simulated point spread functions for a $^{99m}$Tc point source 1 cm off-axis and located at (a) 2cm ,(b) 5 cm and (c) 10 cm from the camera.

**Figure 6.** Point spread functions for the point source located at 12 cm from the collimator, under the cylinder filled with 0 cm, 4 cm and 10 cm water.

**Table 1** summarizes the simulated and experimental FWHM and full width at tenth maximum (FWTM) for the centred and off-centred $^{99m}$Tc point sources.

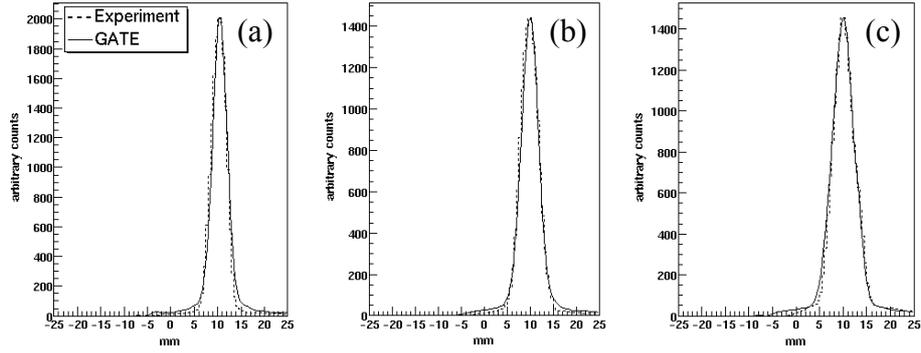

**Figure 5.** Measured and GATE simulated point spread functions for a $^{99m}$Tc point source 1 cm off-axis and located at (a) 2cm ,(b) 5 cm and (c) 10 cm from the camera.

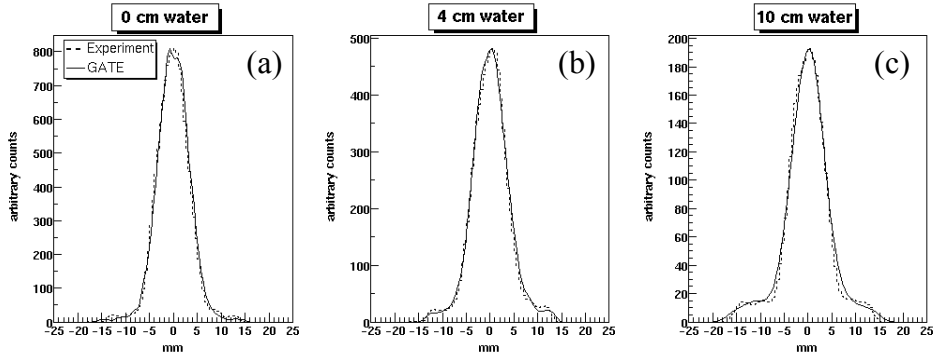

**Figure 6.** Point spread functions for the point source located at 12 cm from the collimator, under the cylinder filled with 0 cm, 4 cm and 10 cm water.

**Table 1**. Comparison between experimental and simulated FWHM and FWTM for the centred and off-centred $^{99m}$Tc sources located at three different distances from the collimator.

| Source-collimator distance (cm) | centred $^{99m}$Tc point source | | | | off-centred $^{99m}$Tc point source | | | |
| | experiment | | simulation | | experiment | | simulation | |
| | FWHM (mm) | FWTM (mm) | FWHM (mm) | FWTM (mm) | FWHM (mm) | FWTM (mm) | FWHM (mm) | FWTM (mm) |
|---|---|---|---|---|---|---|---|---|
| 2 | 3.5 | 5.5 | 3.5 | 6.0 | 3.7 | 6.5 | 3.8 | 6.5 |
| 5 | 4.3 | 8.0 | 4.3 | 8.0 | 4.6 | 8.0 | 4.5 | 8.0 |
| 10 | 6.7 | 12.0 | 6.8 | 12.0 | 5.7 | 9.5 | 5.9 | 11.0 |

Figure 6 shows the measured and simulated point spread functions for the point source located at 12 cm from the collimator, under the cylinder filled with 0 cm, 4 cm and 10 cm water.

## 3.2. Sensitivity

The results of the system sensitivity obtained with GATE compared to experimental measurements are plotted in figure 7 for the five source-collimator distances. Standard deviations obtained by running ten simulations for each distance are also given. Differences between experimental and calculated values were 0.6%, 1.6%, 1.3%, 0.7% and 0.7% for source-to-collimator distances of 2, 5, 10, 15 and 20 cm, respectively.

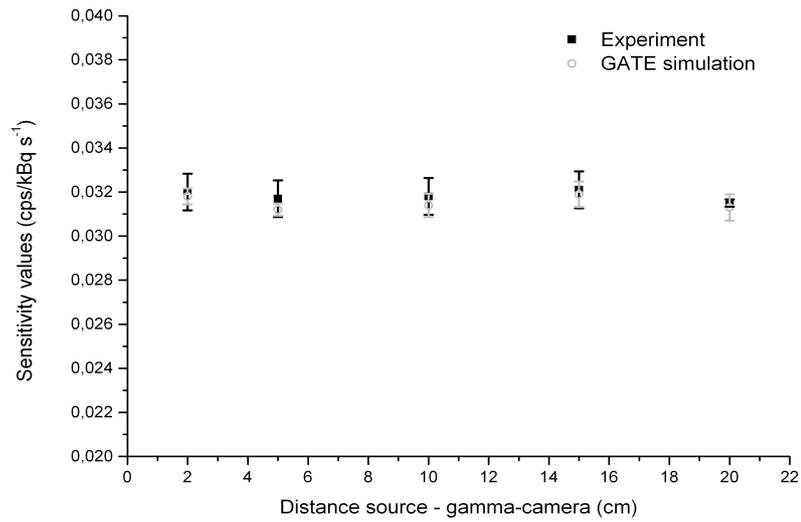

**Figure 7**. Comparison between experimental and simulated system sensitivity for the centred [99m]Tc point source, for 5 distances between the source and the scintillation camera collimator.

**Table 2**. Comparison between experimental and simulated scatter fractions for two source depths and three energy windows.

| Energy window (keV) | 4 cm depth | | | 10 cm depth | | |
|---|---|---|---|---|---|---|
| | Experiment | Simulation | Difference (%) | Experiment | Simulation | Difference (%) |
| 105 – 175 | 0.358 | 0.360 | 0.56 | 0.531 | 0.527 | 0.75 |
| 98 – 182 | 0.380 | 0.379 | 0.26 | 0.539 | 0.546 | 1.30 |
| 91 – 189 | 0.397 | 0.398 | 0.25 | 0.571 | 0.578 | 1.23 |

### 3.3. Scatter fraction

Scatter fractions obtained from experimental measurements and with GATE are given in table 2 for the different energy windows.

The simulated scatter fractions differed by less than 2% from the experimental scatter fractions both in the case of a 4 cm scattering medium and 10 cm scattering medium.

*3.4. Energy spectra*

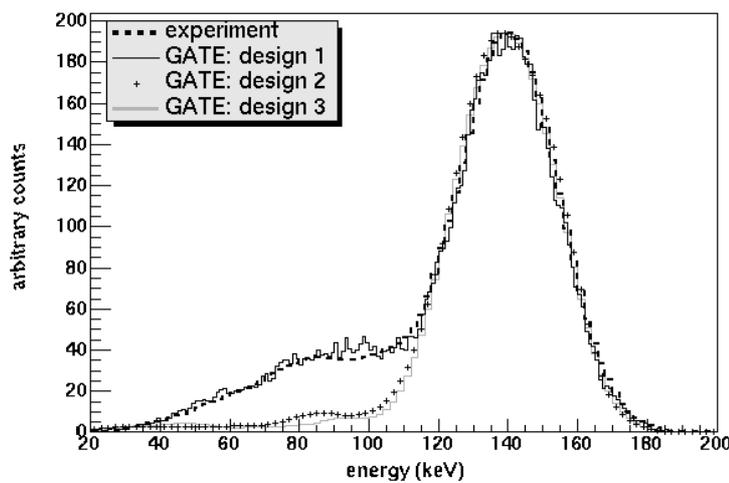

Figure **8** shows the experimental energy spectrum for the $^{99m}$Tc point source and the different energy spectra obtained by simulating different designs of the scintillation camera.

The energy spectra resulting from the simulations of the crystal array (without a PSPMT model behind) with the actual value of the white epoxy gap (design 2: 250 $\mu$m: black crosses) and with a smaller value of 100 $\mu$m (design 3: grey solid line) suggest that this detector component affects the energy spectrum between 70 and 100 keV. This is confirmed by the plot of the contribution of the photons scattered in the PSPMT shown on Figure 9 (left). As demonstrated by the energy spectrum corresponding to design 1 (solid black line), including a PSPMT model was essential to obtain a good agreement between measured and simulated spectra below 110 keV.

Energy spectra obtained for the point source at 12 cm from the collimator, with 0, 4 and 10 cm water thicknesses are shown in figure 9: contributions of photons scattered within different components of the detector (Plexiglas® layer noted PM and Mylar envelope around the pixels) and within the phantom are also plotted. This comparison demonstrates an excellent agreement between simulated and experimental energy spectra.

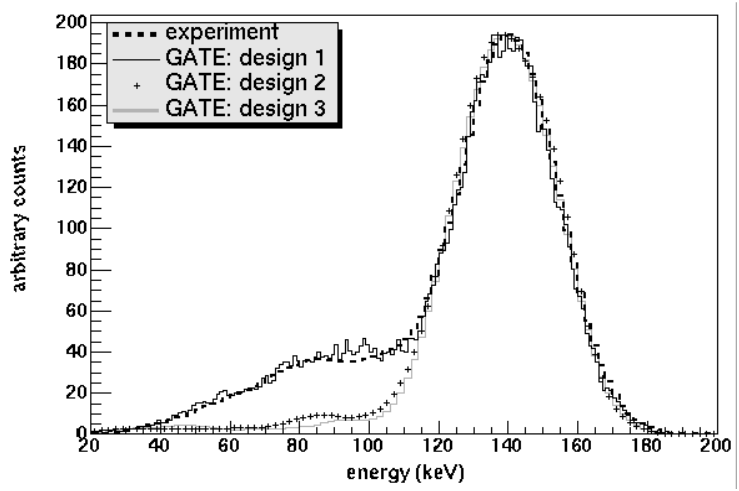

**Figure 8.** Experimental spectrum and spectra simulated using different camera models.

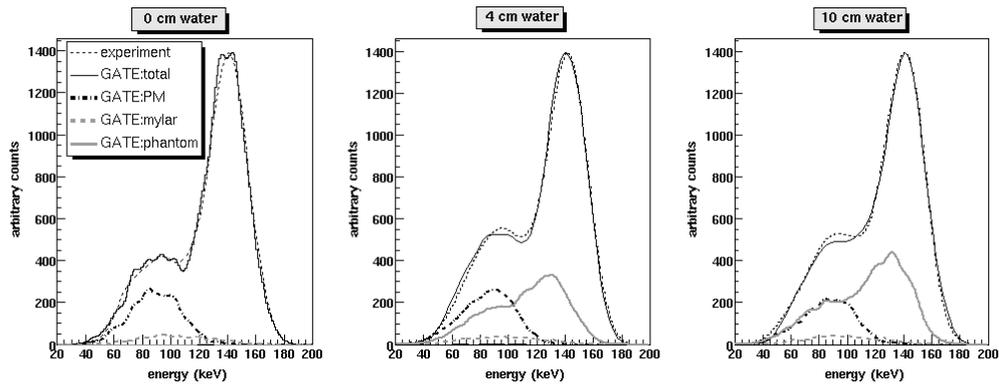

**Figure 9**. Energy spectra obtained for the $^{99m}$Tc point source at 12 cm from the collimator under a water thickness of 0, 4 and 10 cm.

Within the 40-180 energy window, without any scattering medium, the total scatter consisted of 7.8% scatter in the collimator, 16.6% scatter in the Mylar envelope, 74.5% backscatter in the PSPMT and 1.1% in the lead shielding. For the experimental configuration involving 4 cm of water, scatter in the phantom represented 54.3% of total scatter, while scatter in the collimator, Mylar envelope, PSPMT and lead shielding represented 3.5%, 6.9%, 34.8% and 0.5% of total scatter, respectively. For the experimental configuration involving 10 cm of water, scatter in the phantom represented 62.5% of total scatter, while scatter in the collimator, Mylar envelope, PSPMT and lead shielding represented 3.1%, 6.8%, 27.1% and 0.5% of total scatter, respectively.

### 3.5. Image of a capillary phantom

Figure 10 shows the experimental and simulated 64 x 64 images (pixel size = 0.78 mm) of the capillary phantom, as well as profiles (drawn horizontally and centred, 8 pixel thick) through these images with dashed line and solid line, respectively.

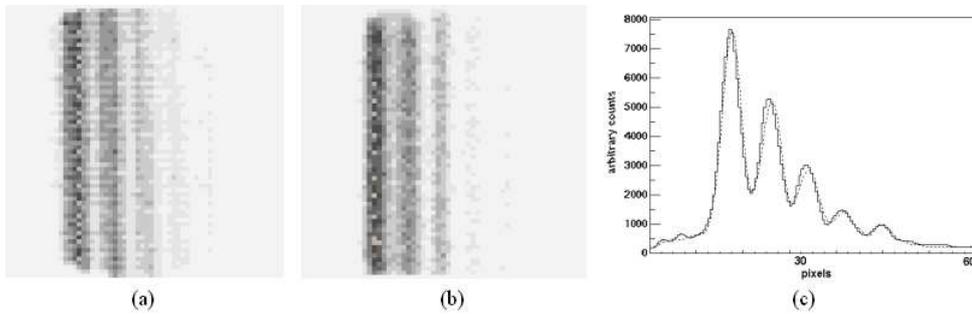

**Figure 10**. Experimental (a) and simulated (b) images of a phantom consisting of 5 capillaries filled with $^{99m}$Tc solutions of different concentrations and horizontal profiles through these images (c: experimental in dashed line and simulated in solid line).

## 4. Discussion

### 4.1. Spatial resolution

Experimental PSF for a centred and an off-centred source were reliably reproduced by the simulations (figures 4, 5 and 6). This suggests that GATE is able to predict the spatial response of the scintillation camera with an accuracy better than 200 $\mu$m if one considers the FWHM values. Discrepancies between simulated and experimental FWTM were lower than 10% in air. Slight differences between simulation and experiments could still be observed in the tails of the PSF: they might be due to imperfect modelling of the PSPMT non-uniform response. Indeed, although the FEP approach globally accounts for non-uniformities introduced by the PSPMT response, local distortions might not be reproduced perfectly using this simple approach.

A comparison between GEANT4 simulations and theoretical calculations of both collimator resolution and sensitivity for parallel-hole and pin-hole collimators has previously demonstrated that GEANT4 enables accurate modelling of interactions within the collimator (Breton *et al* 2001). This feature makes GATE particularly promising for simulating collimator design and configurations involving medium- and high-energy isotope imaging.

### 4.2. Sensitivity

As expected, the experimental sensitivity was constant with distance for the parallel-hole collimator equipping the camera (figure 7). Difference between experimental and simulated sensitivity values were less than 2% for the different source-to-collimator distances that were investigated, demonstrating the reliability of GATE for accurately predicting detector sensitivity.

### 4.3. Scatter fraction

Scatter fractions have been often studied using Monte Carlo simulations for conventional scintillation cameras, but few papers report on them for small-animal scintillation cameras (McElry *et al* 2002).

Simulations using GATE accurately estimated the scatter fractions for a point source 4 cm and 10 cm deep in water, with differences less than 2% and thus GATE can be considered appropriate for estimating the scatter contribution in the scintillation camera under consideration.

### 4.4. Energy spectra

Figures 8 and 9 suggested that, when incorporating measured parameters (here, the FEP map) to model the defects of the detector response, an overall good agreement between simulated and experimental energy spectra could be achieved for various experimental configurations. Some differences could be observed between 90 and 110 keV. These discrepancies might come from the FEP map estimate. The different contributions of scatter (figure 9) show the large proportion of photons which scatter on the PSPMT Plexiglas® layer and confirm the need of modelling this component to properly reproduce the energy spectrum. Photons that scattered in the Mylar envelope rather introduce a small scatter background in the energy spectra.

### 4.5. Image of a capillary phantom

The good agreement between experimental and simulated images and profiles further demonstrate that the simulation was able to closely reproduce the characteristics of experimentally acquired data.

### 4.6. About the GATE simulation platform

This study shows that the high flexibility in geometry description offered by GATE enables the modelling of imaging systems with original design, as well as more conventional systems (Santin *et al* 2003).

GATE uses the cross-section libraries included in GEANT4 (EPDL97, EEDL and EADL), which are at the moment the more up-to-date complete and consistent libraries. The importance of considering appropriate cross-section libraries for scintigraphic imaging simulations has been underlined elsewhere (Zaidi 2000). The good agreement between experimental and simulated data presented in this paper suggests that electromagnetic physical processes are correctly modelled within GEANT4, hence that GATE is reliable from this point of view.

A major drawback of GATE is the computation time required for the simulations, especially because no variance reduction techniques are available in GEANT4 yet. This computation time severely penalizes SPECT simulations as the collimator stops most incident photons, and will even be a greater problem when considering more complex phantom geometries, such as those involving voxelized phantoms. Different strategies are currently under investigation to speed up processing time, among which the parallelization of the code and the 'gridification' of GATE (Breton *et al* 2002). The availability of the GATE simulation platform in the public domain has been announced for June 2004.

## 5. Conclusions

We have shown that GATE enables Monte Carlo simulations of a small-animal imaging scintillation camera prototype. Simulations were found to agree well with experimental measurements in terms of point spread functions, energy spectra, sensitivities, scatter fractions and simple phantom images, suggesting that GATE is appropriate to mimic the behaviour of the prototype. Accurate modelling of the detector response yet requires the incorporation of measured parameters, to account for the defect of the imaging device. Our results also suggest that the energy and spatial responses of the PSPMT and subsequent electronics can be accurately modelled using analytical convolution models. The simulation model will now be used to optimize some of the components of the prototype, to model the response of the scintillation camera in tomographic mode (SPECT) and to study the feasibility of new imaging strategies. As the major drawback of GATE is the computing time required to run the simulations, the implementation of GATE in a computing grid environment is planned to speed up the simulations.

**Acknowledgments**

This work was supported by the French Ministry of Research and Education, the Greek General Secretariat for Research and Technology, under the program PLATON (PAI). We are very grateful to Dr. Christian Morel of the Institute of High Energy Physics (IPHE), University of Lausanne, for initiating the development of the GATE project and for its contribution to this work. This work was partly supported by the Swiss National Foundation for Research under Grant No. 21-63870.00.